\documentclass[
    ,final            
  ]
  {aipproc}

\layoutstyle{6x9}

\usepackage{epsfig}
\usepackage{graphics}
\usepackage{subfigure}
\usepackage{natbib}

\begin{document}

\title[Neutrino emission from GRO J0422+32]
	    {Episodic gamma-ray and neutrino emission from the low mass X-ray binary GRO J0422+32}


\author{Florencia L. Vieyro}{
  address={Instituto Argentino de Radioastronom\'{\i}a (IAR, CCT La Plata, CONICET), C.C.5, (1984) Villa Elisa, Buenos Aires, Argentina}
  ,altaddress={Facultad de Ciencias Astron\'omicas y Geof\'{\i}sicas, Universidad Nacional de La Plata, Paseo del Bosque s/n, 1900, La Plata, Argentina}
}

\author{Yolanda Sestayo}{
  address={Departament d'Astronomia i Meteorologia, Institut de Ci$\rm{\grave{e}}$ncies del Cosmos (ICC), Universitat de Barcelona (IEEC-UB), Mart\'{\i} i Franqu$\rm{\grave{e}}$s 1, E-08028 Barcelona, Spain}
}

\author{Gustavo E. Romero}{
 address={Instituto Argentino de Radioastronom\'{\i}a (IAR, CCT La Plata, CONICET), C.C.5, (1984) Villa Elisa, Buenos Aires, Argentina}
  ,altaddress={Facultad de Ciencias Astron\'omicas y Geof\'{\i}sicas, Universidad Nacional de La Plata, Paseo del Bosque s/n, 1900, La Plata, Argentina}
}  
  \author{Josep M. Paredes}{
  address={Departament d'Astronomia i Meteorologia, Institut de Ci$\rm{\grave{e}}$ncies del Cosmos (ICC), Universitat de Barcelona (IEEC-UB), Mart\'{\i} i Franqu$\rm{\grave{e}}$s 1, E-08028 Barcelona, Spain}
}

\begin{abstract}

GRO J0422+32 is a member of the class of low-mass X-ray binary (LMXB) sources, discovered during an outburst in 1992. Along the entire episode ($\sim 230$ days) a persistent power-law spectral component extending up to $\sim 1$ MeV was observed. These results suggest that non-thermal processes must be at work in the system. We apply a corona model to describe the spectrum of GRO J0422+32 during the flaring phase. We study relativistic particle interactions solving the transport equations for all type of particles self-consistently. We fit the electromagnetic emission during the plateau phase of the event and estimate the emission during an energetic episode, as well as the neutrino production. Our work leads to predictions that can be tested by the new generation of very high energy gamma-ray instruments.

\end{abstract}

\maketitle


\section{Introduction}

The transient source GRO J0422+32 is a member of the class of low-mass X-ray binaries (LMXBs). It was discovered during an outburst in 1992 \citep{paciesas1992}. The episode was also detected in UV/Optical/IR/Radio wavelengths; the data collected in these energy-bands indicate the presence of a black hole in the system. The black hole nature of the compact object is also supported by its spectrum, which shows similarities with other confirmed Galactic black holes, such as Cygnus X-1.

The entire episode lasted about $\sim 200$ days, and in that period a persistent power-law spectral component extending up to $\sim 1$ MeV was observed. This suggests that non-thermal processes should have occurred in the system \citep{ling2003}. In order to explain the non-thermal power-law spectrum of GRO J0422+32 in active state, we propose a model of a magnetized black hole corona \citep{vieyro2012}.

\section{Low-hard state of GRO J0422+32}

At the beggining of the episode, there was a plateau phase that lasted around 15 days, which is identified with the low-had state of the source. In this period, the spectrum can be well described by two components: a thermal component plus a non-thermal tail at higher energies \citep{ling2003}.

To compute the spectral energy distribution (SED), we implement a consistent treatment of non-thermal emission in the magnetized corona. We consider the interaction of locally injected relativistic particles with the matter, photons, and magnetic fields of the corona and the disk, which are taken as background components. This method is based on solving the set of coupled differential equations for all kind of particles through an iterative scheme (for a complete description of the model see the paper by Vieyro \& Romero 2012 \cite{vieyro2012}).

\begin{figure}
  \includegraphics[height=.35\textheight]{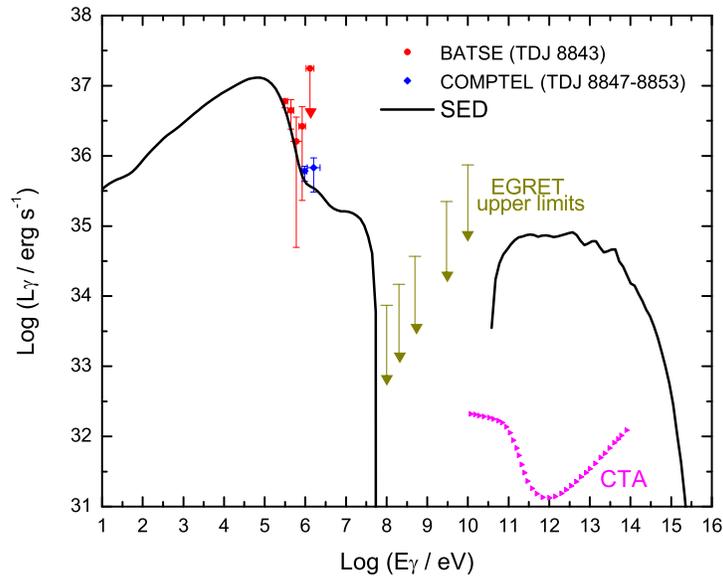}
  \caption{Picture to fixed height}
  \label{fig:steadyLuminosity}
\end{figure}

In Fig. \ref{fig:steadyLuminosity} we show the data of GRO J0422+32 in TJD 8843 during steady state, fitted with the final SED obtained with our model. The best-fit to data is obtained when a 12 \% of the total power available to accelerate particles through magnetic reconnection is injected in relativistic particles.

\section{Flare event}

Within the first 80 days of the 1992 outburst of GRO J0422+32, four shorter but strong episodes were detected in the energy band 0.4 - 1 MeV \citep{ling2003}. The first two of these episodes took place during the plateau phase of the main episode.

We consider non-thermal flares, due to an increment in the power  injected in relativistic particles, and with the thermal corona remaining unaffected during the event. In our model, the power injected in the flare is 15 \% of the luminosity of the corona; this is equal to the total power available for accelerating particles via reconnection events.

\begin{figure}
  \includegraphics[height=.35\textheight]{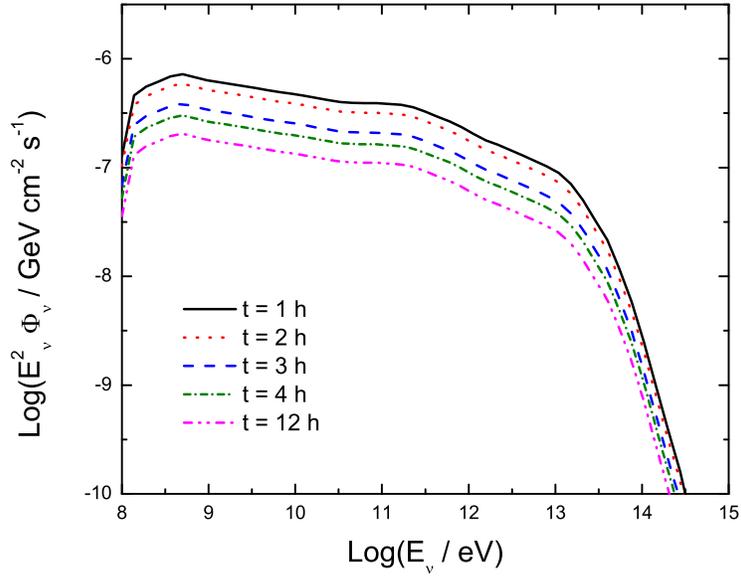}
  \caption{Differential flux of muon neutrinos from a flare in GRO J0422+32 arriving at the Earth.}
  \label{fig:neutrinoFlare}
\end{figure}

It is of particular interest to study neutrino emission during these energetic flares. We consider $\nu_{e}$ production by the channel of muon decay and $\nu_{\mu}$ production by the previous channel plus charged pion decay. In our model the source has null production of initial $\nu_{\tau}$. We take into account the effect of neutrino oscillations to estimate the final flux of $\nu_{\mu}$ arriving at Earth. Figure \ref{fig:neutrinoFlare} shows the evolution of the differential flux of neutrinos at the Earth, weighted by the squared energy, during a flare. 

\section{Detectability of GRO J0422+32 in neutrinos}

We study if the neutrino emission predicted by the corona model during a flare of GRO J0422+32 is detectable by the current high-energy neutrino experiments. We take into account the expected background of atmospheric neutrinos, and instrumental effects such as detection rate and angular resolution. Our calculations for the detection rate make use of the $\nu_{\mu} + \nu_{\bar{\mu}}$ effective area of the IceCube Neutrino Telescope in its 79-string configuration \citep{odrowski2012}.

\begin{figure}
\label{fig:Aeff}  \includegraphics[height=.3\textheight]{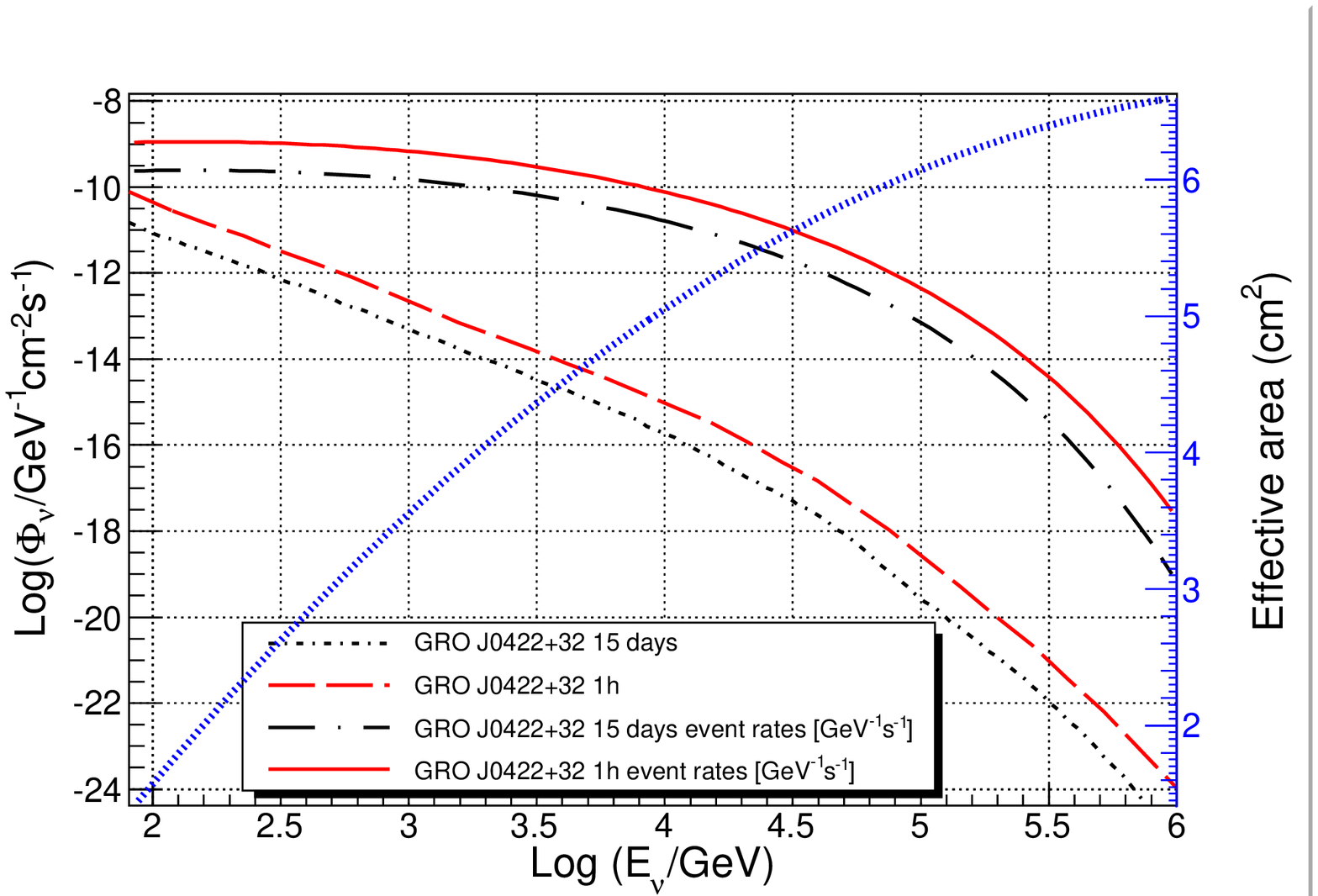}
  \caption{Effective area superimposed to the neutrino spectra of GRO J0422+32 from the 1-hour flare and from the 15-day plateau.}
\end{figure}

Figure \ref{fig:Aeff} shows the effective area superimposed on the neutrino spectra of GRO J0422+32 from the 1-hour flare
and from the 15-day plateau, and the expected event rates as a function of the energy. Above 300 GeV, the total neutrino event rate is $3.62 \times 10^{-6}$ Hz, for the 1h-flare, and $7.72 \times 10^{-7}$ Hz during the 15-day plateau. As a consequence of the energy cutoff at $\sim 8$ TeV, neutrino efficiencies for temporal scales of hours, even days, are too low, and the source, if the corona model is valid, would not be detected.

\section{Discussion}

Under the physical conditions adopted in our model, the main result obtained is that if an outburst with similar characteristic to that observed in 1992 takes place in the present, the probability of detecting it with IceCube is very low. This is because of the very short duration of the high-energy flares observed in this source (less than a day). If the neutrino flux remains as in the peak of the event (Fig. \ref{fig:neutrinoFlare}, t = 1 h) $\sim 80$ days, IceCube will be capable of detecting the  source. It will also be detectable if it remains longer in the plateau phase, for around 690 days.
There are several systems like GRO J0422+32 that show high-energy episodes with the necessary characteristics to enhance the neutrino emission, for example, the outbursts detected from GRO J1719-24, XTE J1118+480 and GX 339-4, among others. Hence, we conclude that longer events in other Galactic sources may be detectable in the future by IceCube.


\begin{theacknowledgments}
 The authors are grateful to James Ling for providing us the data of the BATSE instrument. This work was partially supported by the Argentine Agencies CONICET (PIP 0078) and ANPCyT (PICT 2007-00848), as well as by Spanish Ministerio de Ciencia e Innovación (MICINN) under grant AYA2010-21782-C03-01.

\end{theacknowledgments}



\bibliographystyle{aipproc}   

\bibliography{myrefs_}

\IfFileExists{\jobname.bbl}{}
 {\typeout{}
  \typeout{******************************************}
  \typeout{** Please run "bibtex \jobname" to optain}
  \typeout{** the bibliography and then re-run LaTeX}
  \typeout{** twice to fix the references!}
  \typeout{******************************************}
  \typeout{}
 }

\end{document}